\documentclass[acmtog]{acmart}

\renewcommand\footnotetextcopyrightpermission[1]{} 
\begin{document}

\title{A readahead prefetcher for GPU file system layer}

\author{Vasilis Dimitsas}
\affiliation{%
  \institution{Technion - Israel Institute of Technology}
}
\email{vasileios@alumni.technion.ac.il}

\author{Mark Silberstein}
\affiliation{%
  \institution{Technion - Israel Institute of Technology}
}
\email{mark@ee.technion.ac.il}

\renewcommand{\shortauthors}{V. Dimitsas et al.}

\begin{abstract}
GPUs are broadly used in I/O-intensive big data applications. Prior works demonstrate the benefits of using GPU-side file system layer, GPUfs, to improve the GPU performance and programmability in such workloads. 
However, GPUfs fails to provide high performance for a common I/O pattern where a GPU is used to process a whole data set sequentially. 
In this work, we propose a number of system-level optimizations to improve the performance of GPUfs for such workloads. 
We perform an in-depth analysis of the interplay between the GPU I/O access pattern, CPU-GPU PCIe transfers and SSD storage, and identify the main bottlenecks. 
We propose a new GPU I/O readahead prefetcher and a GPU page cache
replacement mechanism to resolve them. The GPU I/O readahead prefetcher
achieves more than $2\times$ (geometric mean) higher bandwidth in a
series of microbenchmarks compared to the original GPUfs. 
Furthermore, we evaluate the system on 14 applications derived from the
RODINIA, PARBOIL and POLYBENCH benchmark suites. Our prefetching
mechanism improves their execution time by up to 50\% and their I/O
bandwidth by 82\% compared to the traditional CPU-only data transfer techniques.
\end{abstract}

\maketitle

\sloppy

\section{Introduction}

GPUs have been increasingly used in  data-intensive workloads, such as databases~\cite{omniscidb} and graph processing~\cite{gunrock}. Therefore, enabling
efficient GPU access to files and storage devices has become an important optimization goal. 

GPUfs~\cite{Silberstein-gpufs} is the recently introduced system infrastructure which allows GPU threads 
to access files directly from GPU kernels. GPUfs provides standard POSIX-like APIs (e.g.,
({\tt read()}/{\tt write()}) for \emph{GPU threads} to perform file I/O, thus
reducing the programming complexity, it implements a local page cache in the GPU memory, allows
access to very large datasets and enables applications to perform data-driven,
e.g.,  indexed-based, data accesses efficiently. GPUfs passes the I/O requests
to the CPU, yet these remain invisible to the programmer, who only uses a 
standard and convenient  I/O abstraction from GPU kernels. 

While prior works show impressive performance of GPUfs on a range of GPU  workloads, we find
that GPUfs is not as efficient for simple and most common I/O pattern --
streaming-like file access, where  the  whole file is streamed into the
processing unit and sequentially processed chunk-by-chunk. We call this access
\emph{sequential}, because this is how it would be implemented in CPU-only processing.

Sequential pattern is among the most common. For example, it is employed in deep
neural networks~\cite{cudnn} inference and training when reading the input dataset \cite{medium}. 
Such sequential access to data is very popular in high
performance computing. In addition, I/O benchmark suites 
for manycore/multi-node architectures~\cite{ior} consider sequential access as one of the most popular 
in data-intensive applications.

Unfortunately, using GPUfs for sequential accesses results in low
I/O performance. Here, GPUfs API dictates that every GPU threadblock reads one
file stride at a time in a data parallel manner, and then processes that stride
in the threadblock. This access is similar to the one observed in multi-core and
multi-node workloads~\cite{ior}, but results in hundreds of concurrent I/O
accesses to the disk. Our analysis reveals a number of non-trivial performance
problems emerging due to intricate interplay between the GPU I/O access
pattern, the performance characteristics of the PCIe bus for CPU-GPU transfers, 
and the CPU OS readahead prefetching mechanism.  

In this work we show that these limitations are not inherent in GPUfs design,
and demonstrate how to achieve high performance that is singificantly faster
than the original design, and even exceeds the throughput of a commonly-used
traditional CUDA-based \cite{cuda} baseline where files are accessed from the CPU code.
The key to our optimization is \emph{a GPU-side I/O readahead prefetcher and
page cache replacement mechanism}. Together they alleviate the
GPUfs performance limitations, by adjusting the GPU I/O layer to match the
characteristics of the operating system I/O mechanisms and PCIe.

We integrate the prefetcher into GPUfs, and
evaluate it using microbenchmarks and 14 applications derived from the RODINIA \cite{rodinia}, PARBOIL \cite{parboil} and
POLYBENCH \cite{polybench} benchmark suites. We use an NVIDIA K40c GPU with an Intel
P3700 SSD. With the integration of the readahead prefetcher, GPUfs achieves a
$4\times$ (geometric mean) performance than the default GPUfs.
End-to-end, the new GPUfs with readahead prefetcher achieves significant speedups for these applications, e.g.
82\% performance improvement on average (geometric mean) of the GPU I/O sequential read bandwidth compared to CPU I/O. 

The main contributions of this paper are as follows.
\begin{itemize}
    \item Analysis and evaluation of the performance bottlenecks of the CPU-GPU heterogeneous I/O system stack.
    \item Integration of a readahead prefetcher mechanism into GPUfs.
    \item A new cache replacement mechanism for the GPU page cache which enables efficient operation of our GPU readahead prefetcher for files that are bigger than the available page cache size.
    \item Detailed evaluation on microbenchmarks and 14 standard benchmarks from
	    the popular GPU benchmark suites, which show the performance gains
	    of our GPU readahead prefetcher over the default GPUfs and the
	    standard CPU I/O. 
\end{itemize}

\section{Background}
\label{sec:background}

\subsection{System architecture}
We consider a heterogeneous system with a CPU (or a host), a
discrete GPU (or a device) and a fast NVMe SSD, all connected together
through the PCIe bus. The CPU is connected with the host memory, and the discrete GPU is connected with its own physically separate
device memory.

\subsection{GPUfs} 
\begin{figure}[h]
  \centering
 \centerline{\includegraphics[width=1\columnwidth, keepaspectratio]{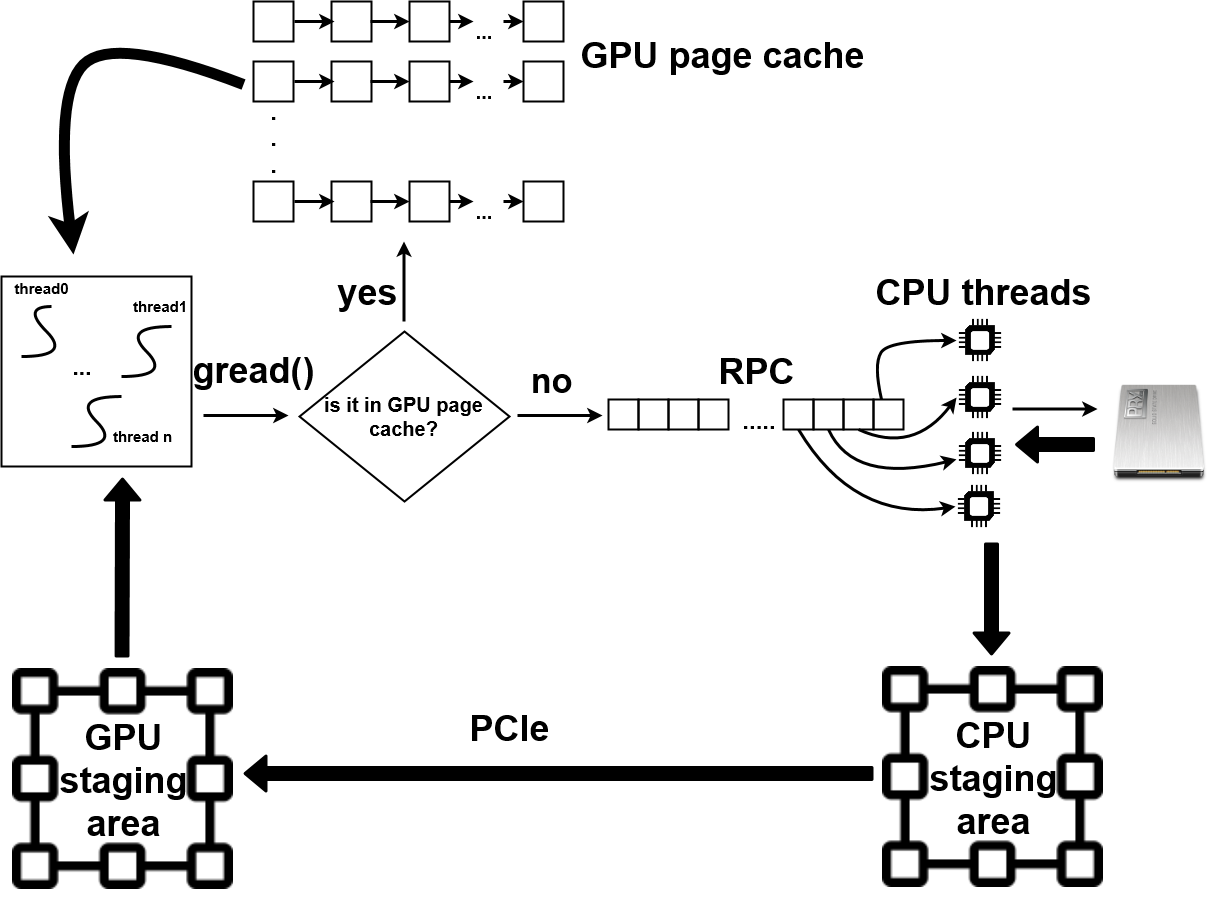}}
	\caption{\label{fig:back:gpufs}The GPUfs file I/O library and its execution flow} 
\end{figure}

GPUfs is a software layer that provides native support for GPU file I/O. The GPU
threads are able to access host files. 
GPUfs provides a
POSIX-like API that is familiar to the programmers and abstracts away the
low-level details of the hardware infrastructure.
 Figure \ref{fig:back:gpufs}
illustrates a high-level view of GPUfs.

Each threadblock may issue an I/O request (read or write) by invoking a file I/O
call from all its threads. The request  may be served from three
locations:  from the SSD (e.g, via peer-to-peer DMA into GPU), the CPU page cache or
in the GPU page cache. For clarity below we discuss only reads, and do not
explain the peer-to-peer I/O because it is beyond the scope of our current
work. See SPIN~\cite{spin} for more details.  

When a threadblock issues an I/O request, GPUfs divides the request into
\emph{pages} stored in the GPU page cache. 
For every page, the GPU page cache is checked and if there is a hit, the data of
the page are copied to the GPU application buffer. In case there is a miss, the threadblock submits 
a request to the shared CPU-GPU request queue (RPC in figure \ref{fig:back:gpufs}) 
and waits for the CPU thread to read the data from disk or the CPU OS page cache. Once
the CPU completes the read, it sends the data to the GPU and signals the threadblock 
that the request has been served and the data is ready for use. After the threadblock receives the 
signal it copies the data to the page  in the GPU page cache and to the user-level buffer.

On the CPU, the host CPU threads poll the shared CPU-GPU request queue, which is
logically divided between them. When a thread finds the request in the queue, it issues the
corresponding {\tt pread()} OS call, one GPUfs page at a time. When the {\tt
pread} completes, the host threads place the data in their staging buffers, and
transfer them to the GPU, batching, opportunistically, multiple chunks together. 

\subsection{Linux Readahead Prefetcher}
A common technique for optimizing I/O performance in sequential accesses is readahead prefetching. 
Linux implements a read-ahead prefetching mechanism that is capable of
optimizing strided sequential accesses in multithreaded programs. 
The
two main features that the Linux Readahead Prefetcher employs are the support of
multiple streams per file descriptor and the asynchronous prefetching operation.
For every file descriptor, there is an allocated readahead data structure which
contains the basic information needed in order to determine the sequentiality of
the read accesses and the prefetch size. 

The prefetcher works as follows. For a given file I/O request it first attempts to
guess whether the application I/O access pattern is sequential, using a few
heuristics. If it is, then the prefetcher issues an asynchronous I/O call of a
larger size than the one requested by the user. In general, the size of the I/O
request is doubled every time the access is considered sequential, until the
size reaches a certain maximum read-ahead threshold.

\textit{Support of multiple strides per file descriptor} 
A common I/O access scenario is when multiple threads access the same file in a
strided manner. Thus, the Linux Readahead Prefetcher applies several heuristics in order to detect the sequential 
access pattern of all the running threads and maintain the efficiency of the readahead prefetcher, even when the threads use the same file descriptor.

\section{Motivation}
\label{sec:motivation}



We run a simple experiment to evaluate the sequential I/O performance of GPUfs
by using it to move 960MB file into GPU memory. The goal is to compare the
effective I/O bandwidth that GPUfs can achieve with the bandwidth
of reading the same file from the CPU (no data transfer to GPU).
We expect that GPUfs would provide full pipe-lining and therefore achieve similar
bandwidth. 

We run the GPU with 120 threadblocks, each 512 threads. Each threadblock reads 8MB stride using
GPUfs. This configuration fully utilizes the GPU compute capacity, and
represents the preferred way to read data into GPUs using GPUfs, as explored in
prior work~\cite{Silberstein-gpufs}. GPUfs is configured with 4KB page size and 4 CPU threads for file system accesses. 
The CPU baseline also uses 4 threads to match the number of CPU threads in
GPUfs.

We observe that the CPU I/O bandwidth is almost 4x higher than the GPU I/O, and reaches 1.6GB/s. 

We thoroughly analyse the reasons for this discrepancy in the next sections.
The goal of our analysis  is two fold: (a) find the root causes of poor GPUfs
performance (b) identify the opportunities for improvements. 

\subsection{Effect of GPUfs page size}
\label{sec:motivation:putting}

\paragraph{Sequential accesses}
One important factor that significantly affects the system performance is the
size of pages in the GPU page cache. That is because GPUfs page size dictates the 
size of accessing files, minimal block transferred over the PCIe, as well as GPUfs page cache management overheads. 
Thus, we seek to find the best GPUfs page size that would enable efficient sequential access. 

\begin{figure}[t]
  \centering
  \includegraphics[width=0.45\textwidth, keepaspectratio]{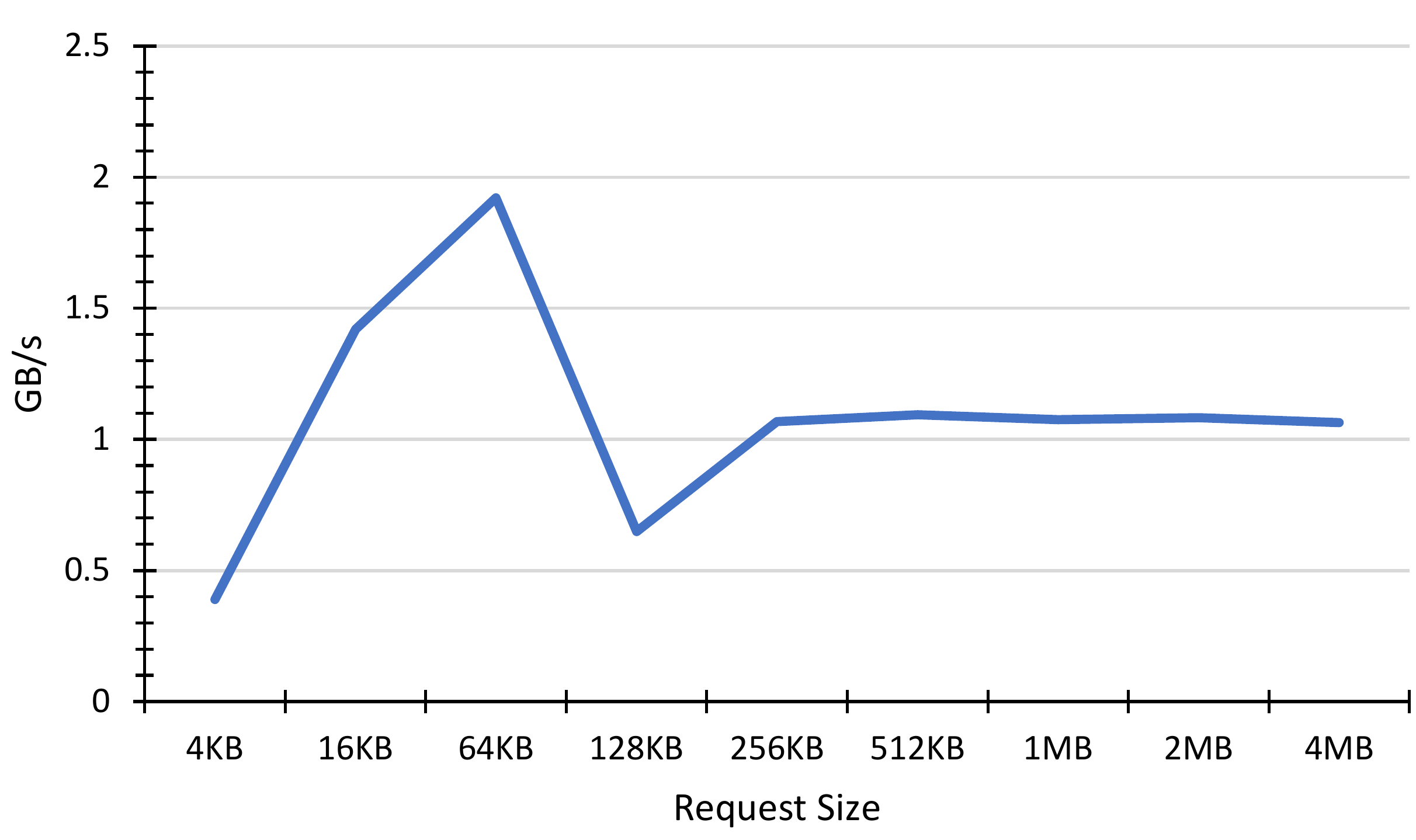}
	\caption{\label{fig:motiv:totalIO}The GPUfs sequential I/O performance
	for different GPU page sizes } 
\end{figure}

Figure \ref{fig:motiv:totalIO} shows that 64KB pages perform the best, and even
exceeding the CPU I/O performance. 

\paragraph{Random accesses}
On the other hand, increasing the page cache page size has negative impact on the
performance of applications with non-sequential accesses that have been shown to
be also important use-cases for GPUfs.

To evaluate the performance impact of the page size in such applications, we run the Mosaic
benchmark~\cite{activep} which creates an image collage from multiple tiny images 
fetched at input-dependent location from a large database (19GB). 
Each tiny image is 4KB. We run the application with GPUfs configured to use 4KB and 64KB pages. 
We observe that smaller pages result in \emph{45\% higher performance} compared to 64KB. 

This is a significant degradation that we seek to avoid. Note that some
applications may exhibit mixed access pattern, whereas some files are accessed
sequentially, and some are not. For example, in the collage application, the
original picture used as the basis for the collage is read sequentially, whereas
the tiny images are accessed in a complex data-dependent manner.

Thus, our analysis below aims to understand the reasons for performance
degradation with 4KB pages, and the rest of the paper relies on this analysis to
overcome this limitation.

\subsection{GPU I/O pattern and the OS I/O layer}
\label{sec:motivation:io}

We first focus on the interaction of GPU-generated I/O pattern with the file I/O layer
on the CPU. Therefore, in this experiment we generate the I/O requests from the GPU
kernel and send them to the CPU, however disable GPU data transfers and
GPU page cache handling.  Thus, we exercise the same GPU access pattern but
without moving the data to the GPU.

Therefore, the only difference between the CPU I/O baseline and the GPU I/O is
the access pattern to storage.

\begin{figure}[t]
  \centering
  \centerline{\includegraphics[width=1\columnwidth, keepaspectratio]{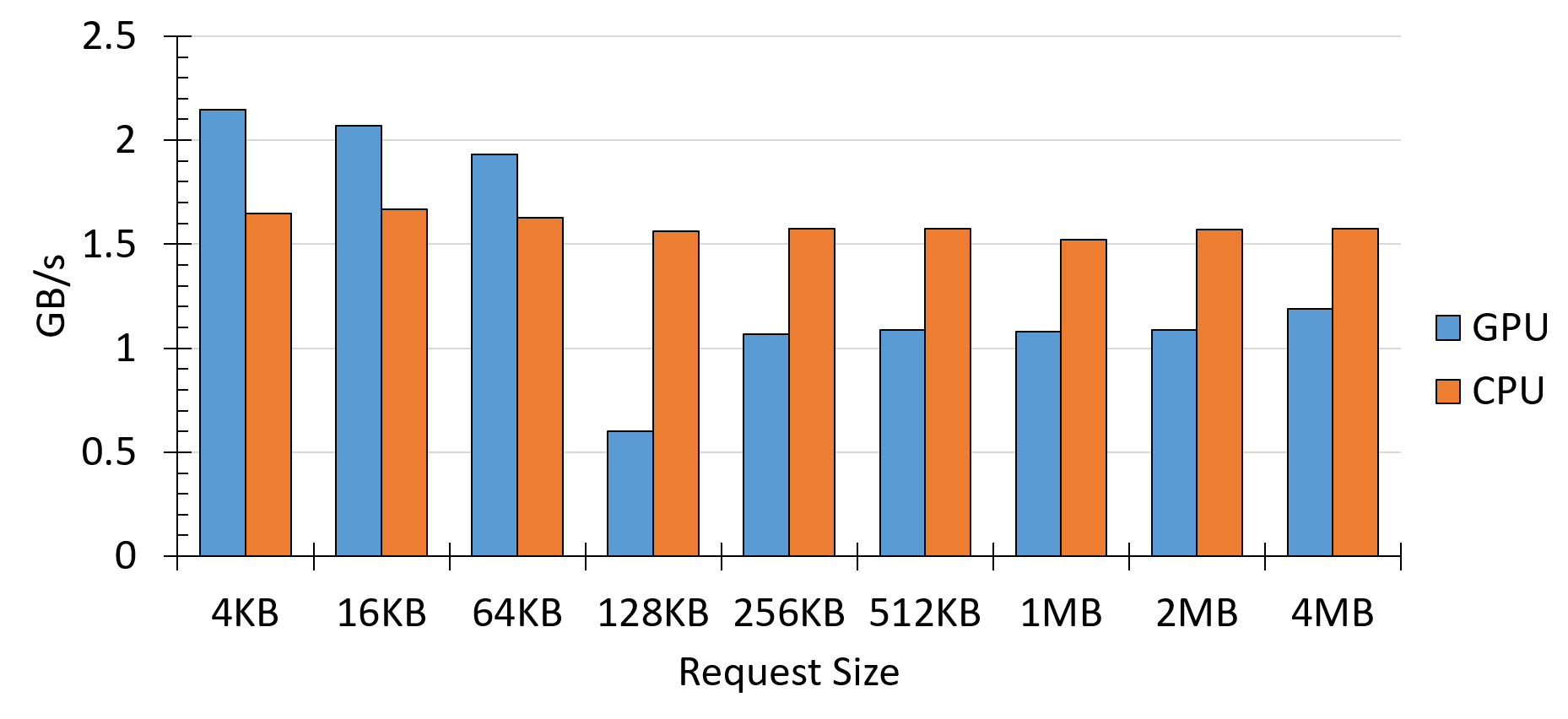}}
	\caption{\label{fig:motiv:GPUvsCPU_onlyIO}GPU vs. CPU I/O bandwidth with
	PCIe transfers disabled.} 
\end{figure}

Figure~\ref{fig:motiv:GPUvsCPU_onlyIO} shows the results. For requests
smaller than 128KB the GPU I/O performance is 24\% better than the CPU
I/O, however for requests above 128KB the CPU I/O bandwidth performance is on
average 61\% higher,  and   160\% higher for 128KB request size.

\begin{figure}[t]
  \centering
  \centerline{\includegraphics[width=.7\columnwidth, keepaspectratio]{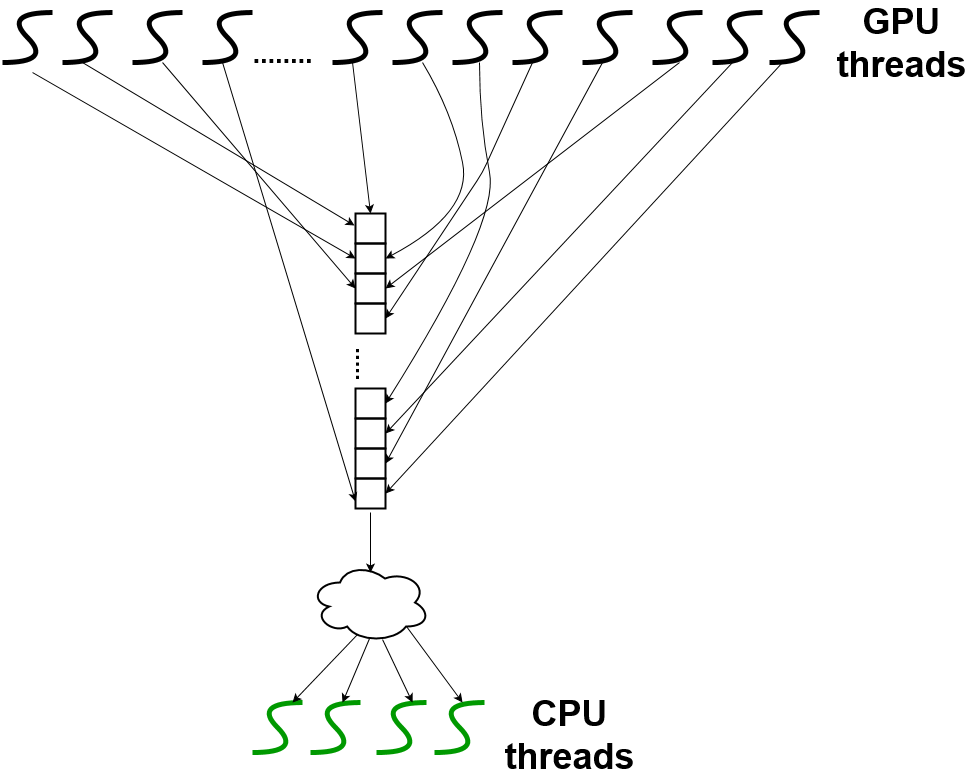}}
	\caption{\label{fig:motiv:gpuCPUWorkMap}The GPU I/O request mapping to the GPUfs host threads.} 
\end{figure}

Figure \ref{fig:motiv:gpuCPUWorkMap} illustrates the I/O request mapping from GPU threadblocks to the GPUfs host threads.
We observe that the host threads serving GPUfs requests observe a random file access pattern. 
This is because the GPU threadblocks are invoked in a non-deterministic order, and thus place their I/O requests in the shared CPU-GPU
request queue in some order that looks random to the CPU threads. 
When the I/O size is below 128KB, the Linux readahead prefetcher manages to
recognize the sequentiality of the GPU I/O accesses and engages the readahead logic that
causes larger reads from the storage. 

Moreover, for smaller reads the GPU I/O pattern enables more efficient use of the prefetched data. 
When a CPU thread issues sequential read requests as in 
Figure~\ref{fig:motiv:gpuCPUWorkMap}, certain requests collide with the request
of the prefetcher, therefore they end up waiting
for data to arrive.  For example, consider the case in which a CPU thread reads 4KB data from
offset 0.  The prefetcher will issue a read I/O request that will fetch 16KB of
data. The thread will be blocked for the first page (4KB) and the rest of the data will be fetched asynchronously. 
If the thread immediately reads additional 4KB (from offset 4K), the
previous asynchronous I/O request by the prefetcher has not yet completed, thus
the thread will block. 

With the I/O pattern  of the GPU execution,  the access to different file portions are interleaved. 
This interleaving enables the prefetcher to issue asynchronous I/O requests
early enough such that by the time the data is accessed it is already in the
page cache. 

On the other hand, for the accesses above or equal to 128KB,  the GPU performance is
much worse. However it cannot be explained by the interaction with the prefetcher, because 128KB is the
maximum size of the read blocks issued by the prefetcher in the default Linux
configuration. Therefore, if the poor performance were due to the prefetcher, it
would remain low for accesses above 128KB, but it is not the case.

\subsection{GPU-CPU interaction}

\begin{figure}[t]
  \centering  \centerline{\includegraphics[width=\columnwidth, keepaspectratio]{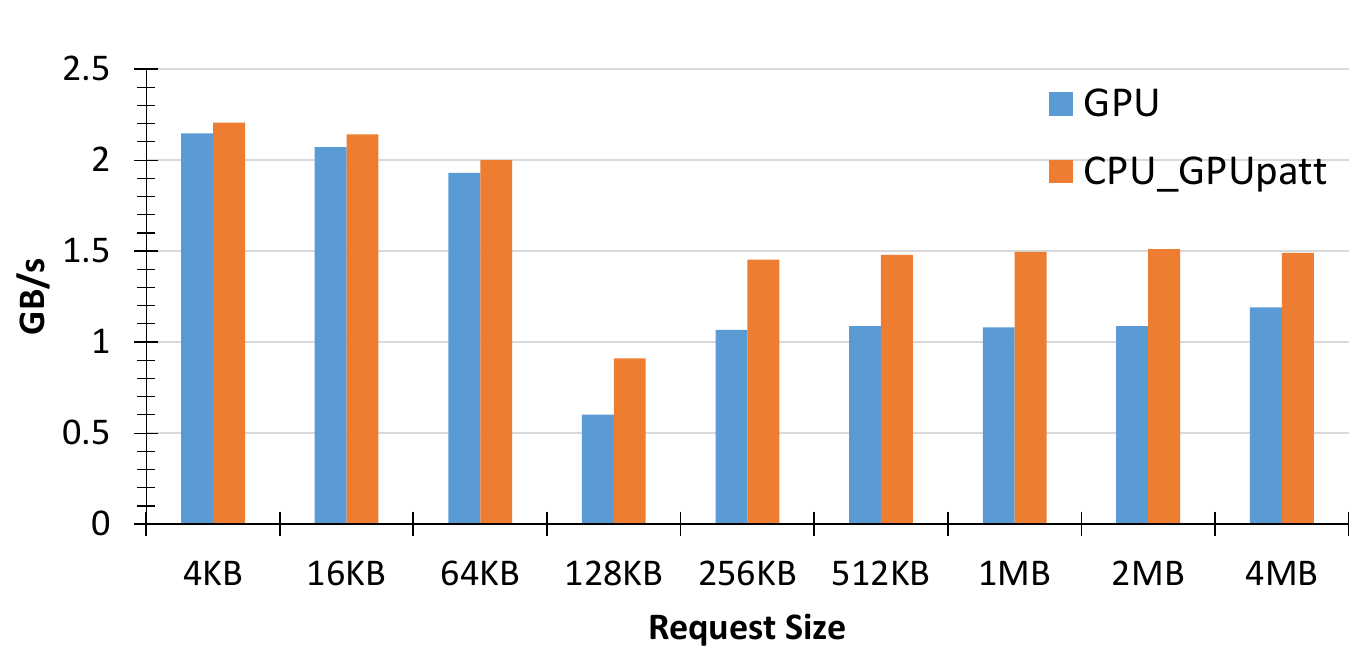}}
	\caption{\label{fig:motiv:GPUvsCPU_onlyIO_GPUpatt}GPU vs. CPU I/O bandwidth performance without PCIe data transfers included. CPU is accessing the file with the GPU I/O access pattern which was observed from the execution of the GPUfs host threads.} 
\end{figure}

To isolate the influence of the file access pattern from the GPU-CPU interaction, 
we record the GPU I/O trace, and then replay it on the CPU without using the GPU. Namely, the
CPU accesses the same file offsets as were accessed by the GPUfs host threads
during GPU I/O. 

Figure \ref{fig:motiv:GPUvsCPU_onlyIO_GPUpatt} shows the difference 
between the GPU I/O bandwidth and the CPU I/O when the latter performs the same I/O
pattern. 
As expected, the performance is nearly the same for accesses below 128KB. 

However, for the requests of 128KB and more, we observe that the performance of
the GPU-initiated accesses is much worse. The main reason lies in the CPU-GPU
interaction through the shared I/O queue. The four CPU threads serving GPU I/O
requests periodically scan their own portions of the queue, but it turns out that
some of them remain \emph{idle} because of the poor load balancing, and this
idling occurs only for requests equal or grater than 128KB.

\begin{figure}[t]
  \centering
  \includegraphics[width=1\columnwidth, keepaspectratio]{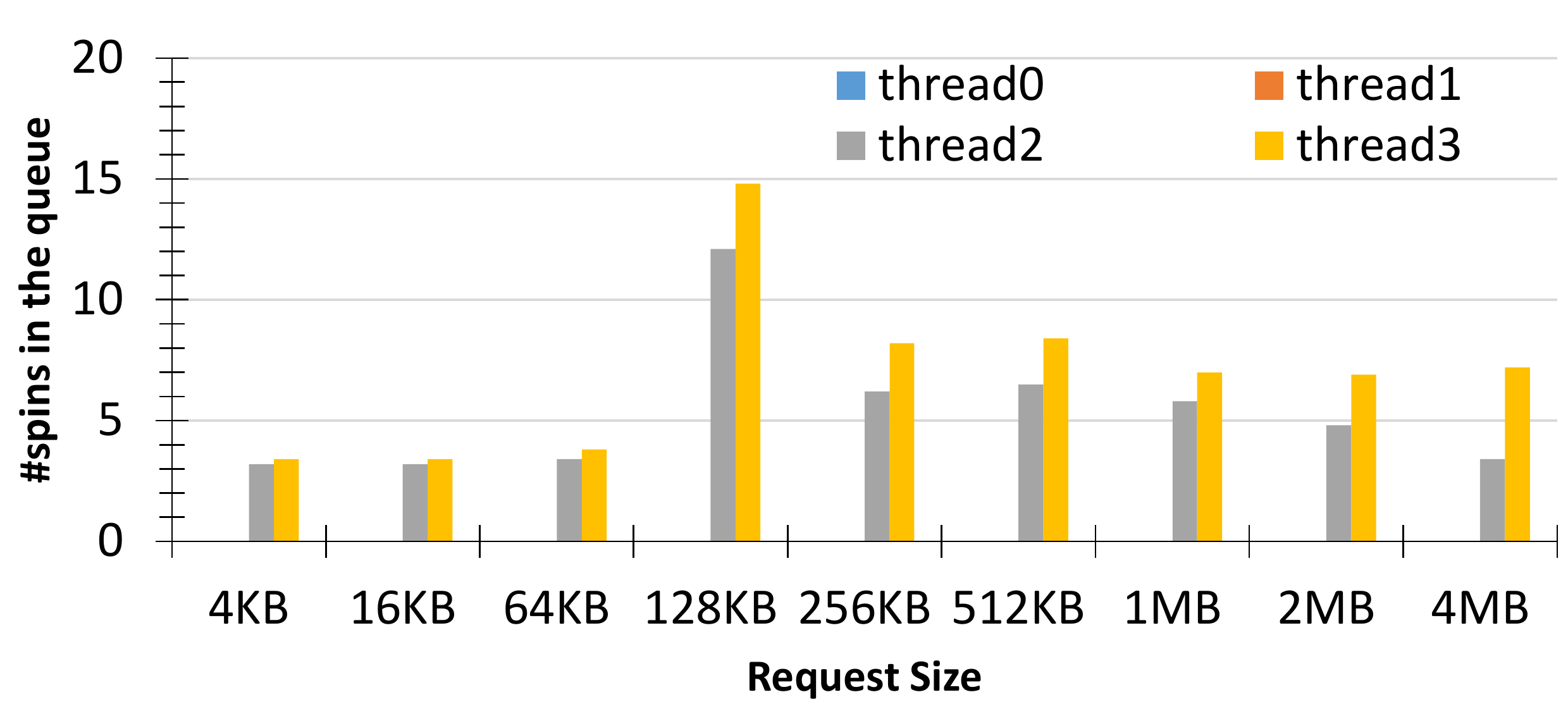}
	\caption{\label{fig:motiv:spins}The number of spins the GPUfs host threads make before they start servicing their first I/O requests} 
\end{figure}

Figure \ref{fig:motiv:spins} shows the waiting time (in
millions of polling attempts) that each GPUfs host thread spends before it starts servicing its first request. We observe that threads 0 and 1 are busy
(their bars are invisible), whereas threads 2 and 3 are idle for larger pages.

The reason lies in the way GPUfs GPU threads assign work requests. When a
threadblock issues an I/O request, it assigns the request to the slot of the
shared queue based on its CUDA threadblock ID, to avoid contention for the queue
slots. The shared queue has 128 slots, where each CPU thread polls its set of
contiguous 32 slots. 

When the GPU is invoked with 120 threadblocks in our experiments, 
only 60 threadblocks fit the GPU at the same time (15SMs, maximum 2048 threads
per SM). Therefore, in the beginning of the run the GPU I/O requests are served only by two CPU
threads. 
The phenomenon affects more the requests with size equal or greater than 128KB. The reason lies in the I/O performance of the corresponding requests. For 64KB or less, the requests are executed faster and thus, the waiting time of the idle threads is less compared to the bigger requests.

\subsection{PCIe data transfers}
\label{sec:motivation:pci}
\begin{figure}[h]
  \centering
  \includegraphics[width=\columnwidth, keepaspectratio]{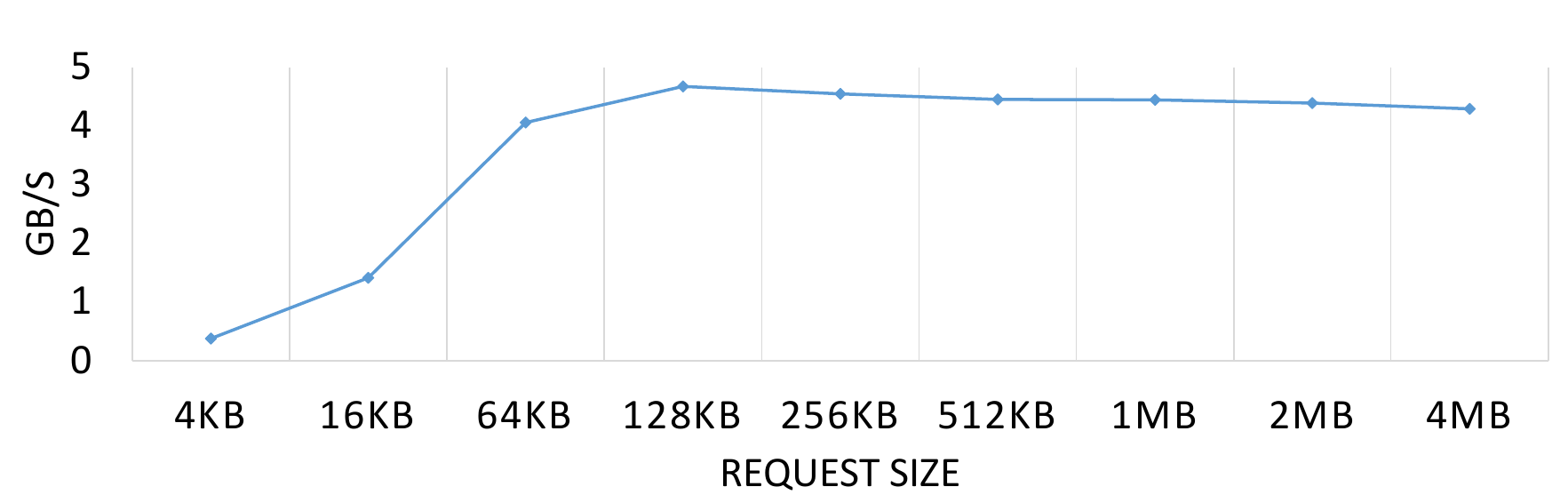}
	\caption{\label{fig:motiv:pcie}The PCIe bandwidth performance of for the GPU microbenchmark} 
\end{figure}

To isolate the impact of PCIe data transfers from the file I/O, we perform the
same experiment as above while the data is stored in memory in RAMfs \cite{ramfs}.
Figure \ref{fig:motiv:pcie} shows, as expected, that larger page sizes perform
much better than the small ones. This is in conflict with earlier
observations that smaller pages benefit the system performance.

\subsection{Conclusions.} 
We seek to improve GPUfs performance with 4KB pages, and observe that the main bottleneck is the PCIe bus. Thus, the main optimization opportunity is to increase the size of PCIe transfers by \emph{prefetching} the data in advance into the GPU page cache in larger chunks.

\section{The GPU I/O Readahead Prefetcher}
\label{sec:GPUIO_design}

\label{sec:design_considerations}

\paragraph{GPU I/O readahead prefetcher in the CPU?} 
The location of the prefetch logic is an important design question. One option
is to place the prefetcher on the CPU side. Thus, for every GPU I/O request,
the CPU might move larger blocks to the CPU. This design, however, requires the
CPU to be aware of the internal memory allocation in the GPU file system layer
and the page cache, and is hard to implement without having PCIe support for
atomic operations.  Our design uses GPUfs itself to request larger chunks from
the CPU, thereby allowing the on-GPU GPUfs logic to allocate memory, and ensure
consistent access to data. This design is also similar to the way the OS
prefetcher is implemented in the CPU.

\paragraph{Asynchronous operation.}
Performing prefetching asynchronously is a natural design choice. However, it is
hard to combine it with the GPUfs PCIe transfer layer. In GPUfs, when the CPU
completes the GPU read request, it transfers the data into a staging buffer in
the GPU memory, and each threadblock retrieves the page that fulfills its own
request as part of the GPU I/O call. The purpose of the GPU prefetcher is to move
data to the GPU in larger chunks which include more pages in addition to the one
in the request. Thus, by the time the GPU threadblock retrieves the requested
page, the prefetched data is already in the GPU, hence the benefits of
asynchronous prefetching are diminished. Thus, our design uses synchronous prefetching. 

\subsection{Design}
\begin{figure}[h]
  \centering
  \includegraphics[width=0.4\columnwidth, keepaspectratio]{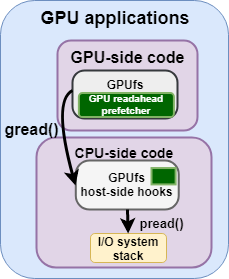}
	\caption{\label{fig:design:gpuIOsystem}The GPU file I/O system stack} 
\end{figure}

Figure \ref{fig:design:gpuIOsystem} shows the main components of the GPU I/O system stack. 
The GPU I/O readahead prefetcher is part of GPUfs and it is primarily located in
the GPU. 

In this section we illustrate the interaction of the GPU I/O
readahead prefetcher with {\tt gread()} call in GPUfs.

\subsubsection{GPU-side mechanism}

\begin{enumerate}
    \item Every threadblock has its own private buffer for holding the prefetched data.
    \item When a threadblock performs a gread(), it accesses the GPU page cache.
    \item If the data is there, it is used to fulfill the read request and the
	    GPU execution continues.
    \item If there is a page cache miss, the threadblock allocates a new page
	    and searches if the data is available in its private buffer.
    \item If the data is in its private buffer, it updates the page cache, returns the data to the user-level buffer and continues execution.
    \item If the data is not found in the private buffer, then the threadblock
	    issues a read request to the CPU with size equal to GPUfs PAGE\_SIZE + PREFETCH\_SIZE (PREFETCH\_SIZE is static and is defined before execution).
    \item When the data is transferred back from the CPU, the threadblock
	    updates the page cache with the 4KB page that is currently requested
	    and the rest of the pages are placed in its private buffer.
\end{enumerate}

The CPU modifications required for the integration are mainly to enable the CPU handle the management of the multiple GPUfs page cache pages. When the corresponding pread finishes its execution, it returns the actual size of the data that has been loaded. The CPU thread, therefore, will examine how many GPUfs pages will be delivered to the GPU. The main operations are setting their metadata (e.g. size, offset per page) and placing the data in the staging buffer.

\noindent{\bf Why per-threadblock private buffers?}
The original version of GPUfs enables file accesses at threadblock-level
granularity. Therefore per-threadblock private buffers made the design, implementation and integration of the GPU I/O 
readahead prefetcher more efficient. Additionally, the private buffer is a reasonable solution
because it is quite likely that the same threadblock will access the prefetched
pages. Last, having a private buffer eliminates the page cache contention among
the threadblocks, thereby reducing the overhead of dealing with prefetching. In
fact, our original design used the page cache as a prefetcher target instead of
the private buffers, but it suffered from significant synchronization overheads.

\noindent{\bf Lack of a global prefetching scheme.}
The GPU I/O readahead prefetcher does not implement any synchronization
mechanisms between the threadblocks for coordinated prefetching. 
Therefore, there is a possibility that different threadblocks may prefetch the
same page  if the access pattern is not sequential.  

However, coordinating the access among the threadblocks would incur significant performance penalty, as it would
require synchronization between the threadblocks accessing the same page. 
Therefore, the user could disable in such cases the GPU I/O readahead prefetcher by using runtime hints like {\tt
posix\_fadvise()}~\cite{fadvise} that is used for the CPU I/O readahead prefetcher.

\noindent{\bf Warp-level prefetching.}
The newest version of GPUfs \cite{data-driv-gpufs} enables the GPU threads to perform disk
I/O at warp-level granularity. In that case, the GPU I/O readahead prefetcher
design would have been modified. Warps should synchronize in order to access the
shared per-threadblock private buffers or assign in each warp its own private
buffer. Furthermore, a warp-level granularity prefetcher should be designed by
taking into consideration that there would be needed less aggressive prefetching
because the rate of concurrent I/O requests is higher. We do not address
warp-level I/O in this work.

\noindent{\bf Page cache coherency.}
The prefetcher mechanism that relies on private buffers works correctly for
read-only workloads. However, when the file is modified, this solution might
result in a stale copy of the page in the private buffer. Specifically,  if a given page is 
retrieved by multiple threadblocks at the same time (thus multiple copies reside
in the private buffers of each accessing threadblock), then modified in the page cache
by one of the threadblocks, and then evicted from the page cache, the 
copy still remaining in the private buffer of some threadblock will be wrong. 

One way to solve this problem is to maintain a global bit map of dirty pages in
each file, updating it when pages get modified by the GPU.
The respective bit is checked before accessing the private buffer in stage (5),
discarding the locally-cached version if the page was modified. 

Our current implementation does not include this additional mechanism,
however, because the vast majority of GPU workloads are read-only. 
Therefore, we enable prefetching for files opened in read-only mode, and disable it for
writable files, allowing to optimize performance for the most common cases. We
defer the complete implementation for future work.

\section{Large files}

Sequential access to files that exceed the (configurable) size of the GPUfs page
cache results in severe thrashing of the GPU page cache and dramatic performance
degradation (See the baseline performance in Figure~\ref{fig:eval:bigFiles}).
Note that the use of 4KB pages amplifies the effects of the thrashing
significantly, and cannot enjoy the benefits of prefetching. The straightforward solution of increasing the GPU page cache is not efficient due to the limited size of the GPU memory. Thus, it would not be a scalable solution that could be applicable for every case.

\subsection{A new page cache replacement mechanism}

Our new page cache replacement mechanism is optimized to avoid page cache
management overheads for large files that are read sequentially.
Our main idea is that each threadblock keeps its \emph{own} set of active
pages. Assuming that when the data is accessed sequentially it will be used only
once, each threadblock effectively performs local replacement mechanism. 

More specifically, each threadblock keeps its own Least Recently Allocated (LRA) queue of pages it
allocates in the GPU page cache when reading a file. 
The queue is of a fixed size which is determined as the total size of the page cache divided by the
number of actively concurrently running threadblocks (i.e., the number of threadblocks that
reach maximum GPU occupancy.) Whenever the threadblock allocates a new page in
the page cache, it adds the page to its list. When it runs out of pages, it
evicts the least recently allocated page in its private list, and places there a
new page,  while remapping the page as necessary in the page table and
re-inserting it into the head of the Least Recently Allocated queue. 

This mechanism entirely eliminates the synchronization overheads incurred in
managing a global LRA queue shared among all the threadblocks, and does not
require a page to be de-allocated and allocated again -- which is how it is
implemented in the original GPUfs implementation.

\section{Evaluation}

\begin{table*}[t]
	
\begin{center}
\begin{tabular}{|l|l|l|}
	\hline
	\bf BENCHMARK  & \bf NUMBER OF INPUT FILES AND SIZES & \bf I/O KERNEL
	CONFIGURATION \\
   \hline
	HOTSPOT & 2 files, 1 GB each & 128 tblocks, 512 threads \\
   \hline
   LUD & 1 file, 256 MB & 128 tblocks, 512 threads \\
   \hline
   BACKPROP & total data read 3.25 GB & 128 tblocks, 512 threads \\
   \hline
   BFS & 1 file, 1.1 GB & 128 tblocks, 512 threads \\
    \hline
   DWT2D & 1 file, 768 MB & 128 tblocks, 512 threads \\
   \hline
   NW & 2 files, almost 1 GB each & 100 tblocks, 512 threads \\
   \hline
   PATHFINDER & 2 files, almost 1 MB and 952 MB & 100 tblocks, 512 threads \\
   \hline
   STENCIL & 1 file, 1 GB & 128 tblocks, 512 threads \\
   \hline
   2DCONV & 1 file, 1 GB & 128 tblocks, 512 threads \\
   \hline
   3DCONV & 1 file, 512 MB & 128 tblocks, 512 threads \\
   \hline
   GESUMMV & 1 files, almost 1 GB & 128 tblocks, 512 threads \\
   \hline
   MVT & 1 file, almost 1 GB & 128 tblocks, 512 threads \\
   \hline
   BICG & 1 file, almost 1 GB & 128 tblocks, 512 threads \\
   \hline
   ATAX & 1 file, almost 1 GB & 128 tblocks, 512 threads \\
   \hline
 \end{tabular}

\end{center}
\caption{\label{tbl:eval:benches}The benchmarks which are used for the evaluation of the GPU I/O readahead prefetcher and the new GPU page cache replacement mechanism (tblocks is an abbreviation for threadblocks).}
\end{table*}

We implement our design into GPUfs and evaluate it on NVIDIA Tesla K40C GPUs,
Intel Xeon E5-2620v2 CPU with 64GB DDR4 DRAM, and Intel NVMe SSD DC P3700
with 2.8GB/s read bandwidth. We evaluate it on Ubuntu 15.04 with Linux kernel
3.19.0-47 and ext4 on SSD. We use CUDA 6.5 for GPUfs, the GPU programs and the
GPU microbenchmarks. Finally, we run every experiment 10 times and we calculate the average value (arithmetic mean). 
We flush the contents of the CPU page cache before every experiment.

\begin{figure}[h!]
  \centering
 \centerline{\includegraphics[width=.8\columnwidth, keepaspectratio]{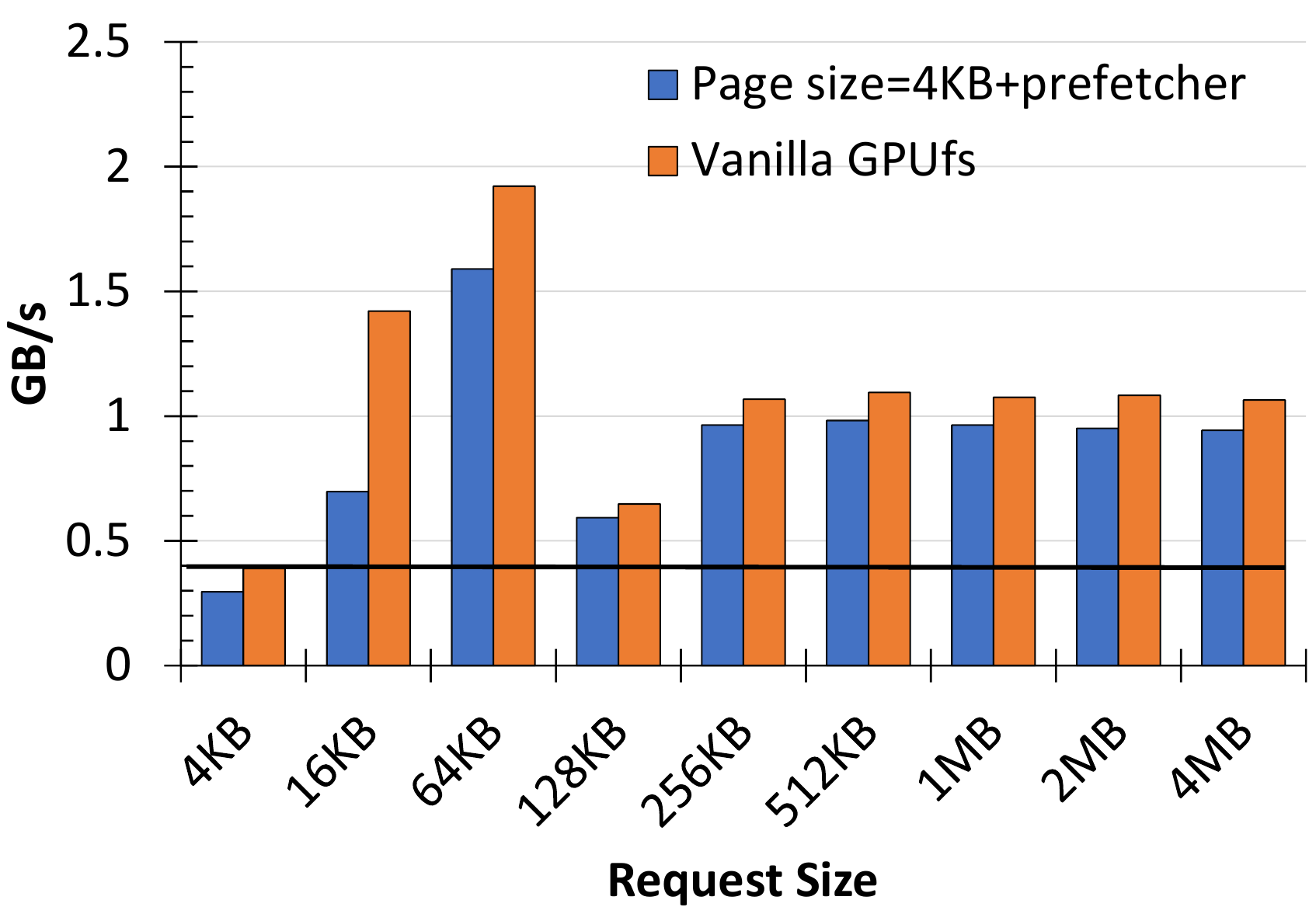}}
	\caption{\label{fig:eval:totalPerf}The GPU I/O performance when GPUfs
	uses the GPU I/O readahead prefetcher with 4KB pages, vs. the original version of GPUfs
	with large pages} 
\end{figure}

\begin{figure}[h!]
  \centering 
  \centerline{\includegraphics[width=.9\columnwidth, keepaspectratio]{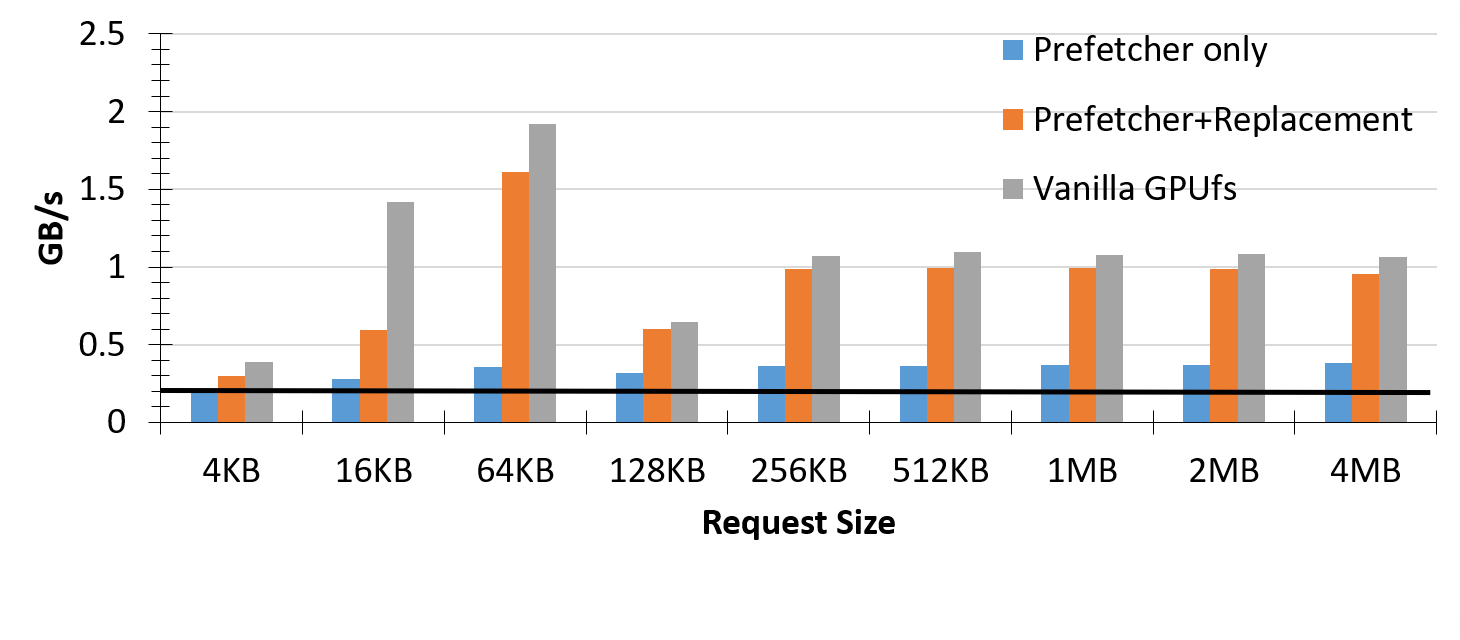}}
	\caption{\label{fig:eval:bigFiles}The GPU I/O performance of the
	microbenchmark for files that do not fit in the GPU page cache. We
	compare GPUfs with the new page cache replacement mechanism and the
	prefetcher vs. the GPUfs with the prefetcher only vs. GPUfs original version} 
\end{figure}

\subsection{Microbenchmarks}
The microbenchmark we use for our evaluation is a GPU application configured
with 120 threadblocks, 512 threads per threadblock. The application reads 1 GB
from a 10GB file.  Every threadblock issues the same number of sequential I/O
read requests into its own strides of the file in a data-parallel manner.

\paragraph{The performance of the GPU I/O readahead prefetcher}
We compare the benefits of the GPU prefetcher over the original GPUfs as
follows. We vary the page size from 4KB to 4MB in the original GPUfs. In the new
version that uses the prefetcher we vary the \emph{prefetch} size accordingly,
but keep the GPUfs page size 4K. In this experiment we expect that the new GPU
prefetcher will help recover the GPUfs performance with large pages, but without
actually modifying the size of the GPUfs page.

Figure \ref{fig:eval:totalPerf} shows that the prefetcher achieves significant
improvement, within 20\% of the best performance of GPUfs with 64KB pages. This
is about $2\times$ better than the original GPUfs.

\paragraph{Big files.}
We evaluate our new page cache replacement mechanism by executing the same
microbenchmark but now reading 4GB (twice the size of the page cache) vs. 1GB in
the previous experiment. 
Figure \ref{fig:eval:bigFiles} shows that the use of the new replacement mechanism
is significantly better compared to the use of 4KB pages without it (blue
baseline). 

\subsection{Benchmark applications}
\label{par:GPUIO15}
We evaluate the GPUfs with the new readahead prefetcher and page cache
replacement mechanism on 14 benchmarks derived from the RODINIA, PARBOIL and POLYBENCH benchmark suites. 
We follow the methodology used in prior works to evaluate GPU I/O
performance~\cite{nvmmu} as follows. The input for the GPU kernel in the original
benchmarks is first stored in a file, and then the benchmarks are invoked when the data is read from the file into GPU memory. The
benchmark time measurement has been modified to include the time to read the
file, copy it to GPU memory, and run the GPU kernel. 

We scale up the file sizes either by using the tools provided with the
benchmark, or by generating large files ourselves.
Table \ref{tbl:eval:benches} presents the benchmarks and their I/O configuration. The reason for configuring the execution of NW and PATHFINDER at 100 threadblocks is because of the file size that these specific benchmarks process. Thus, we set the number at 100 so that the threadblocks process equally-sized portions of the file. 

In the experiments we compare three implementations: (1) CPU I/O -- standard CPU I/O (1 CPU
thread) that reads the whole file at once with  cudaMemcpy to transfer data to the GPU; (2) 
GPUfs with 64KB pages to show the upper bound on the performance as it results in
the best performance; (3) GPUfs with 4KB pages and the prefetcher configured to
64KB. (4) The original GPUfs with 4KB pages without the prefetcher. GPUfs is configured with
2GB page cache in all of the cases.

\paragraph{File size smaller than the page cache size}
 In this experiment we evaluate the performance when the page cache is large enough to store the entire input.

\begin{figure*}[t]
  \centering
  \centerline{\includegraphics[width=1\textwidth, keepaspectratio]{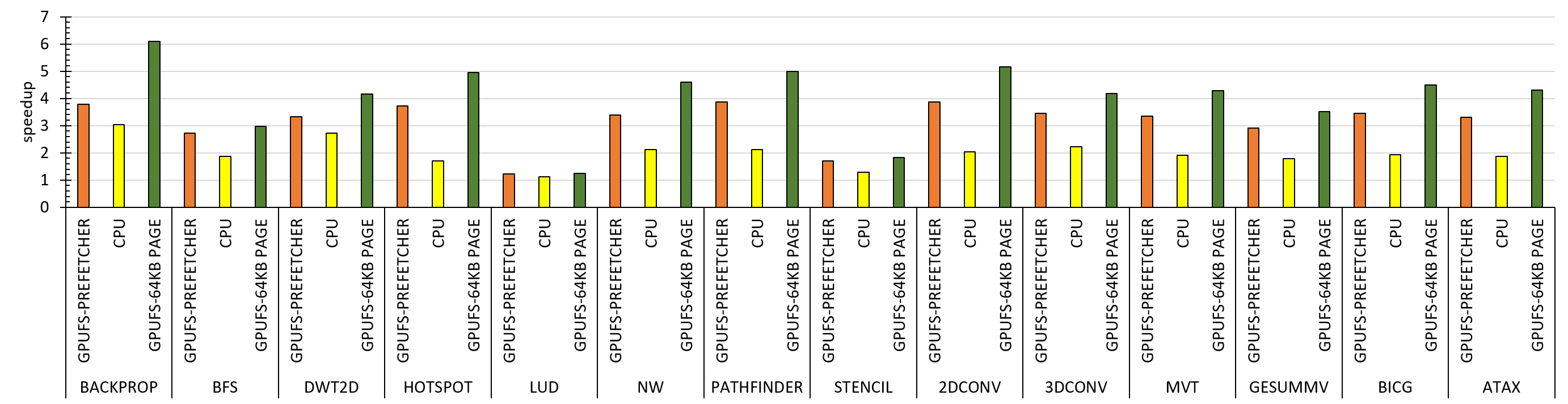}}
	\caption{\label{fig:eval:totalExecRatio15} End-to-end speedup over
	original GPUfs with 4KB GPUfs pages: 
	GPUfs-prefetcher -- this work with 4KB GPUfs pages, CPU -- CPU-only (original benchmark),
	GPUfs-64KB -- GPUfs with 64KB pages. Files are smaller than the page cache.}
\end{figure*}

Figure \ref{fig:eval:totalExecRatio15} shows the end-to-end speedup for the
benchmarks over the unmodified GPUfs with 4KB pages.

The use of GPUfs prefetcher improves the performance  of the original GPUfs by  
$3\times$ on average (geometric mean), compared to the default version of GPUfs,
and by over $1.5\times$ on average (geometric mean) compared to the CPU I/O.

\begin{figure*}[t]
  \centering
  \centerline{\includegraphics[width=1\textwidth, keepaspectratio]{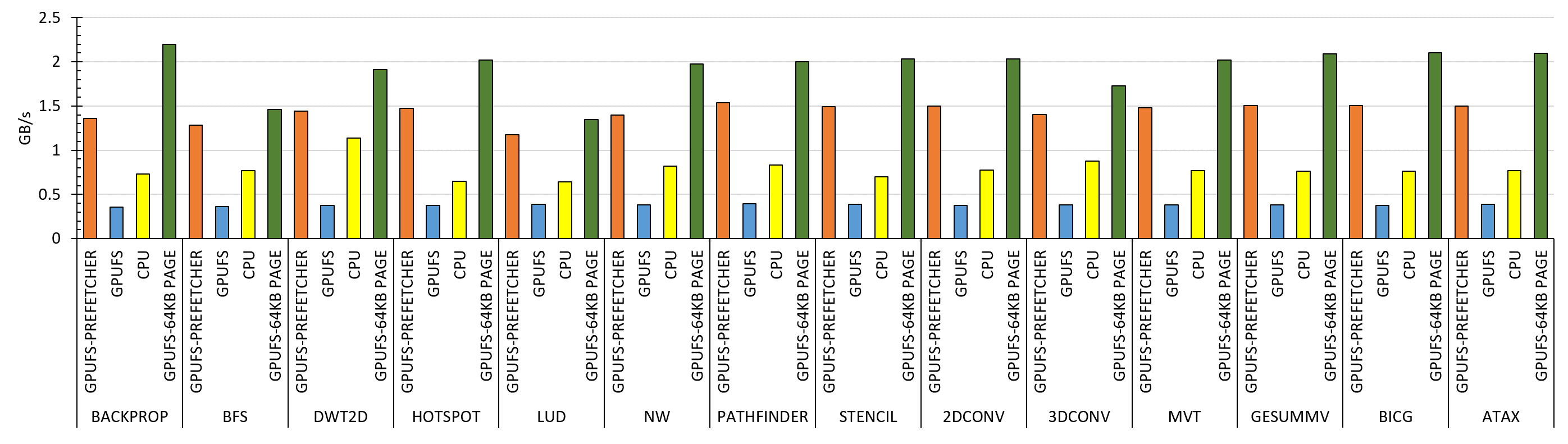}}
	\caption{\label{fig:eval:IOband15}
	I/O performance of: 
	original GPUfs with 4KB pages, 
	GPUfs-prefetcher -- this work with 4KB GPUfs pages, 
	CPU -- CPU-only (original benchmark),
	GPUfs-64KB -- GPUfs with 64KB pages. Files are smaller than the page cache.}
\end{figure*}

To show that this speedup stems from the improved I/O performance, Figure
\ref{fig:eval:IOband15} shows the I/O bandwidth for each benchmark.

The results indicate that the speedup indeed originates from the efficient I/O.
More specifically, the I/O bandwidth with the prefetcher is on average (geometric mean) almost $4\times$ higher  
than with the original GPUfs version and $2\times$ higher than with
the CPU I/O.

\begin{figure*}[t]
  \centering
  \centerline{\includegraphics[width=1\textwidth, keepaspectratio]{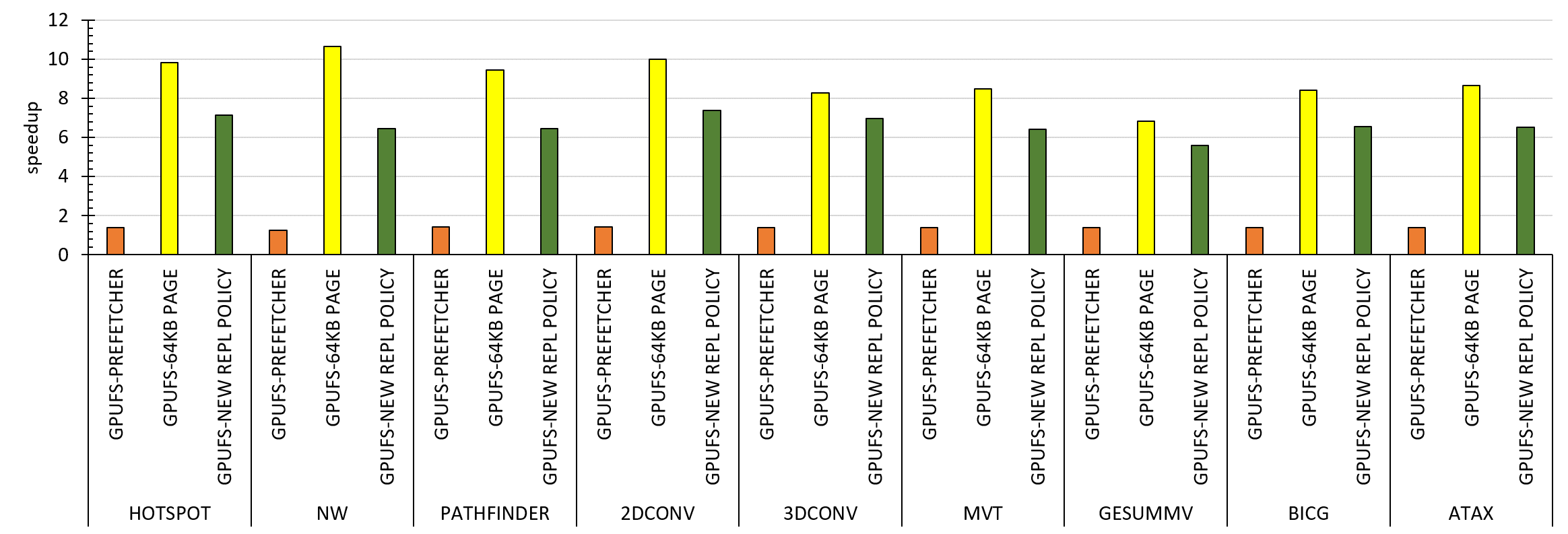}}
	\caption{\label{fig:eval:totalExecRatio9_big} End-to-end speedup over
	original GPUfs with 4KB pages:
	GPUfs-prefetcher -- this work with 4KB GPUfs pages without the new
	replacement policy, GPUfs-64KB -- GPUfs with 64KB pages, GPUfs-new
	replacement -- this work with the new replacement policy. Files are
	\emph{larger} than the page cache.}
\end{figure*}

\begin{figure*}[t]
  \centering
  \centerline{\includegraphics[width=1\textwidth, keepaspectratio]{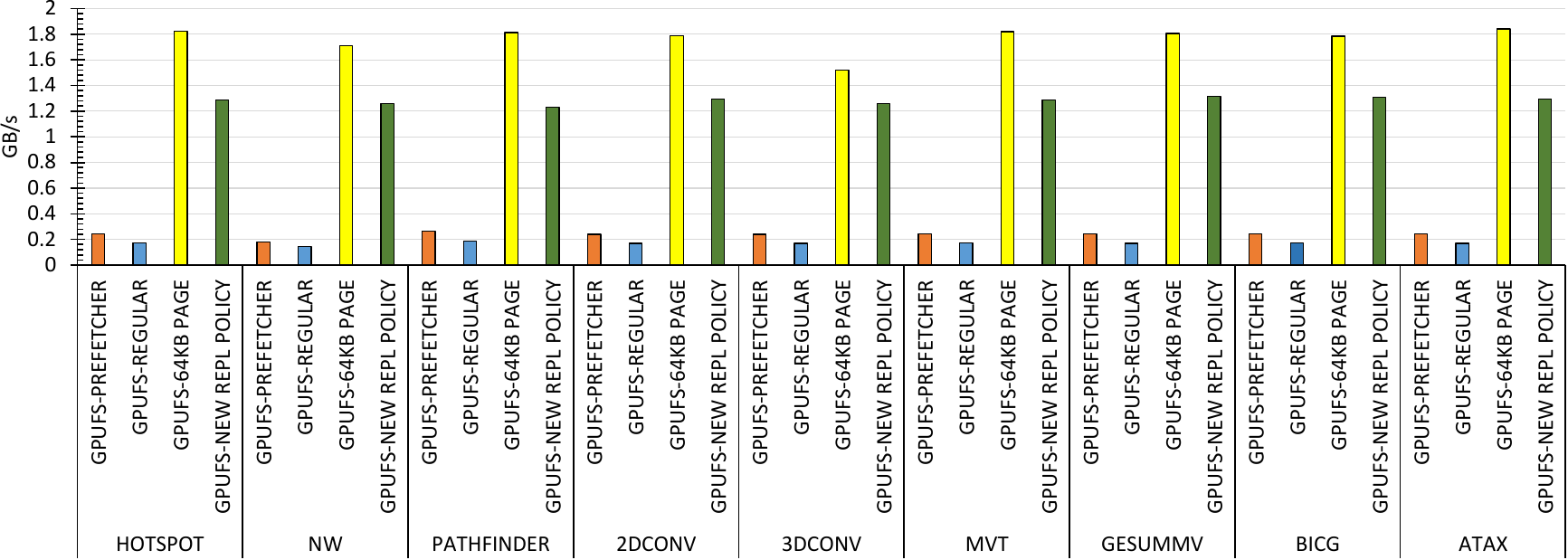}}
	\caption{\label{fig:eval:IOband9_big}
	I/O performance of:
	GPUfs-prefetcher -- this work with 4KB GPUfs pages without the new
	replacement policy, 
	GPUfs-regular -- original GPUfs with 4KB pages,
	GPUfs-64KB -- GPUfs with 64KB pages, GPUfs-new
	replacement -- this work with the new replacement policy. Files
	are \emph{larger} than the page cache.
	}

\end{figure*}

\paragraph{File size larger than the page cache size}
We perform the same experiments,  but this  time we configure the GPU page cache
to be smaller than the total input size. Specifically, we configure it to be 500MB 
(except for setting the page cache size to 256MB for 3DCONV because the input data size is 512MB).

We compare three versions: GPUfs with 4KB pages with the enabled GPU prefetcher
with 64K and new replacement mechanism, original GPUfs with 64KB pages (best
performing configuration for GPUfs) and GPUfs with the prefetcher but without
the new replacement mechanism.

Figure~\ref{fig:eval:totalExecRatio9_big} shows the end-to-end speedup over GPUfs with 4KB
pages. The average speedup (geomean) due to the new replacement mechanism is almost
$5\times$.

Figure~\ref{fig:eval:IOband9_big} shows the performance improvement to the I/O bandwidth, when GPUfs uses the prefetcher along with the new page cache replacement mechanism.
It is almost $6\times$ more on average (geometric mean) than the version of
GPUfs that uses only the prefetcher and $8\times$ (geometric mean) higher against the default version of GPUfs.

\paragraph{GPU I/O prefetcher vs bigger GPU page cache page sizes}
All the new mechanisms introduced in this paper enable faster operation of GPUfs
for 4KB pages in case of sequential accesses. However, we 
observe that configuring GPUfs with 64KB pages achieves higher performance. 
This is expected, as the use of 4KB pages in GPUfs incurs higher page cache management
overheads compared to managing 64KB pages, and this paper does not deal with
optimizing the page cache data structures.



\section{Related work}
\label{sec:related}

Data prefetching has been extensively studied and used for many cases. To the best of our knowledge, it is the first time that a system-level, I/O readahead prefetcher is proposed and integrated with the GPU I/O system stack. In this section we present a subset of the related work that exists in bibliography.

\textbf{Hardware prefetchers.} Many hardware prefetching schemes have been proposed by prior work such as \cite{stateless}\cite{tlb-pref}\cite{data-pref}. \cite{apogee} is a mechanism which dynamically detects and adapts to the graphics and scientific applications memory access patterns. \cite{graph-gpu} introduces a prefetching technique that boosts the performance of graph applications which have been implemented in GPU. \cite{compass} introduces a prefetching mechanism for CPUs that takes advantage of the idle shaders of the integrated GPUs and improves the single-threaded performance. \cite{many-thread-gpgpu} consists of a hardware and software prefetcher tailored to GPGPU systems but focus mainly on enhancing the performance of the memory accesses. On the contrary, our GPU I/O readahead prefetcher aims to improve the performance of the GPU I/O sequential reads, regardless of the application type, is a software component and is integrated with the GPU I/O system stack.

\textbf{I/O prefechers.} \cite{pref-cache-opt} introduces a new algorithm that makes the minimum number of I/O instructions to service the I/O requests. \cite{mpi-parall-pref} proposes an I/O signature-based prefetcher which mainly targets to detect the I/O pattern of an application and issues the requests as early as possible. \cite{aggr-pref} raises the issue of performing more aggressive I/O prefetching in terms of speculation and data size. Our work follows this direction but also considers other factors as well, like the PCIe data transfers, because of the complexity of a heterogeneous CPU-discrete GPU system. \cite{hide-IO} pre-executes a fragment of code, via a pre-execution thread that runs at the same time as the main thread, in order to prefetch I/O requests. This solution cannot be efficiently applied for GPUs because it would require code divergence and would raise performance issues.

\textbf{System level I/O prefetchers.} \cite{advOS} is a system level I/O readahead prefetcher, which is integrated with the Linux OS and enhances the I/O performance of the programs which perform sequential I/O read accesses. As our study showed, the Linux Readahead Prefetcher is not sufficient to enhance the I/O performance of the GPUs because of the diverse architectural characteristics of a heterogeneous CPU-discrete GPU system. The GPU I/O readahead prefetcher operates synergistically with the Linux Readahead Prefetcher in order to improve the GPU I/O performance.

\textbf{GPU system abstractions.} \cite{Silberstein-gpufs} and \cite{gpunet} provide file access and networking abstractions to GPU programs in order to be able to issue I/O and networking instructions without the intervention of the CPU. The current work is complementary to GPUfs as it is integrated into it and enhances the performance of the GPU applications with I/O sequential read access pattern.

\section{Conclusion}
\label{sec:conclusion}

We introduce a GPU I/O readahead prefetcher which enhances the performance of the applications that perform per-threadblock sequential I/O read accesses in discrete portions of a file. We make a thorough analysis of the GPU I/O system stack components in order to identify the bottlenecks and examine the I/O characteristics of the GPU applications that could potentially harm or improve their performance. We show that the I/O access pattern does not affect the performance for request sizes greater than or equal to 128KB and for smaller request sizes, it actually boosts their performance because the address interleaving offers efficient overlapping for the asynchronous I/O prefetch requests, which are issued by the Linux Readahead Prefetcher.

We observe and conclude that the GPUfs workload scheduler, which maps the GPU I/O requests to the host threads, creates a detrimental effect on the GPU I/O performance, for request sizes greater than or equal to 128KB. Furthermore, the GPU memory limitations also cause performance degradation for applications that operate on files that do not fit in the GPU page cache and thus, we introduce a new page cache replacement mechanism that reduces significantly the number of allocations/deallocations of pages and therefore, the data structure handling overheads.

The new computing paradigm dictates the need for OS support for the heterogeneous systems \cite{omnix} and the expansion of the legacy OS abstractions in order to provide greater flexibility and convenience to the heterogeneous system programmers. The GPU I/O readahead prefetcher is a step towards this direction as it expands the efficiency of the Linux Readahead Prefetcher for GPU I/O.

\bibliographystyle{ACM-Reference-Format}
\bibliography{paper.bib}
\end{document}